\documentclass[twocolumn,superscriptaddress,aps,showpacs,amsmath,footinbib,bibnotes]{revtex4-1}
\pdfoutput=1
\usepackage{graphicx}
\usepackage{amssymb,amsmath}
\usepackage{textcomp} 
\usepackage[bold]{hhtensor}
\usepackage{natbib}
\usepackage{lipsum}
\usepackage{bm}
\usepackage{hyperref}
\usepackage{multirow}
\usepackage{float}
\usepackage{amsmath}
\usepackage{amssymb}
\usepackage{stmaryrd}
\usepackage{mathrsfs}
\usepackage[dvipsnames]{xcolor}
\hypersetup{
	colorlinks,
	linkcolor={red!80!black},
	citecolor={Blue},
	urlcolor={MidnightBlue}
}

\renewcommand \d{\mathrm{d}}
\renewcommand \i{\mathrm{i}}

\newcommand \ad{\dagger}

\begin{document}

\author{Mian Zhang}
\thanks{These authors contributed equally to this work}
\affiliation{John A. Paulson School of Engineering and Applied Sciences, Harvard University, Cambridge, Massachusetts 02138, USA}
\author{Cheng Wang}
\thanks{These authors contributed equally to this work}
\affiliation{John A. Paulson School of Engineering and Applied Sciences, Harvard University, Cambridge, Massachusetts 02138, USA}
\affiliation{Department of Electronic Engineering, City University of Hong Kong, Kowloon, Hong Kong, China}
\author{Yaowen Hu}
\affiliation{John A. Paulson School of Engineering and Applied Sciences, Harvard University, Cambridge, Massachusetts 02138, USA}
\affiliation{Department of Physics, Tsinghua University, Beijing, 100084, China}

\author{Amirhassan Shams-Ansari}
\affiliation{John A. Paulson School of Engineering and Applied Sciences, Harvard University, Cambridge, Massachusetts 02138, USA}
\affiliation{Department of Electrical Engineering and Computer Science, Howard University, Washington DC 20059, USA}
\author{Tianhao Ren}
\affiliation{John A. Paulson School of Engineering and Applied Sciences, Harvard University, Cambridge, Massachusetts 02138, USA}
\affiliation{University of Electronic Science and Technology of China, Chengdu, 611731, China}
\author{Shanhui Fan}
\affiliation{Ginzton Laboratory and Department of Electrical Engineering, Stanford University, Palo Alto, California 94305, USA}
\author{Marko Loncar}\email{loncar@seas.harvard.edu}
\affiliation{John A. Paulson School of Engineering and Applied Sciences, Harvard University, Cambridge, Massachusetts 02138, USA}

\date{\today}
\title{Electronically Programmable Photonic Molecule}

\begin{abstract}
Physical systems with discrete energy levels are ubiquitous in nature and are fundamental building blocks of quantum technology. Realizing controllable artificial atom- and molecule-like systems for light would allow for coherent and dynamic control of the frequency, amplitude and phase of photons. In this work, we demonstrate a photonic molecule with two distinct energy-levels and control it by external microwave excitation. We show signature two-level dynamics including microwave induced photonic Autler-Townes splitting, Stark shift, Rabi oscillation and Ramsey interference. Leveraging the coherent control of optical energy, we show on-demand photon storage and retrieval in optical microresonators by reconfiguring the photonic molecule into a bright-dark mode pair. These results of dynamic control of light in a programmable and scalable electro-optic platform open doors to applications in microwave photonic signal processing, quantum photonics in the frequency domain, optical computing concepts and simulations of complex physical systems.
\end{abstract}
\maketitle

Photonic analogues of condensed matter systems have resulted in important discoveries like photonic crystals \cite{joannopoulos_photonic_1997}, parity-time symmetric systems \cite{regensburger_paritytime_2012} and topological photonic systems \cite{lin_photonic_2016,lu_topological_2014}, and have led to breakthrough technologies including quantum ground state cooling of nanomechanical systems \cite{chan_laser_2011}, new classes of sensors \cite{hodaei_enhanced_2017,chen_exceptional_2017} and one-way lasers \cite{bandres_topological_2018}.A photonic analogue of a two-level system could allow full-control over the energy and phase of photons using the concept of two-level systems control in atomic or molecular systems, where the state of the electron can be controlled and functionalized by external electromagnetic fields. Such a photonic system would enable the investigation of complex physical phenomena \cite{joannopoulos_photonic_1997,regensburger_paritytime_2012,lu_topological_2014,bandres_topological_2018,kues_-chip_2017} and unique functionalities, including on-demand photon storage and retrieval, coherent optical frequency shift and optical quantum information processing at room temperature \cite{kues_-chip_2017,obrien_optical_2007,liu_observation_2001}. However, the fundamental building block, a coherent photonic two-level system that can be programmed and dynamically controlled, is still missing. While realizing a photonic device with discrete energy levels is straightforward, for example using modes of an optical resonator, controlling such a system dynamically is challenging as it requires mixing of optical frequencies via strong nonlinear processes. As a result, coherent coupling between discrete photon energy modes have only been studied using all optical methods \cite{guo_-chip_2016,ramelow_strong_2018,sato_strong_2012}, which require high optical pump powers and have limited design parameter space, configurability and scalability.

\begin{figure*}
	\centering
	\includegraphics[angle=0,width=\textwidth]{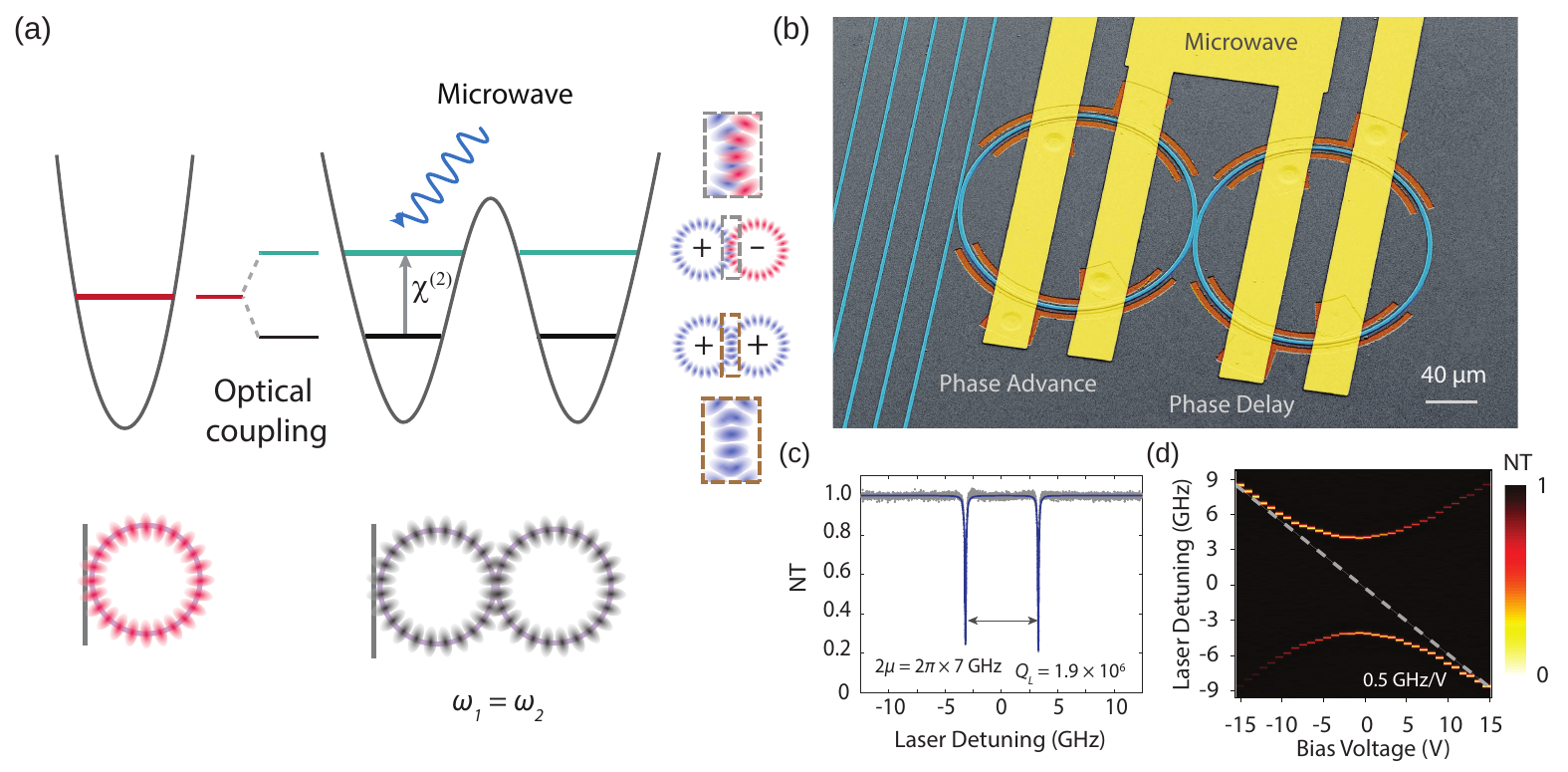}
	
	\caption{\label{fig1}\textbf{Microwave-controlled photonic molecule.} (a) The photonic two-level system is realized by a pair of identical coupled optical microring resonators ($\omega_1 = \omega_2$). The coupling forms a photonic molecule with two distinct energy levels, a symmetric (S) and an antisymmetric (AS) mode that are spatially $\pi$ out of phase. Microwave field can interact coherently with the photonic two-level system through the strong Pockels effect enabled by second order nonlinear ($\chi^{(2)}$) of lithium niobate. (b) False-coloured scanning electron microscope (SEM) image of the coupled microring resonators fabricated on thin-film lithium niobate. To maximize the effective microwave-photonic interaction strength, gold microwave electrodes are patterned to achieve differential driving that induces a phase delay on one ring and phase advance on the other one. (c) Measured transmission spectrum of the photonic two-level system. The two optical modes are separated by $2\mu = 2\pi \times 7$ GHz with linewidths of $\gamma = 2\pi \times 96$ MHz, which corresponds to a loaded (intrinsic) optical quality factor of $1.9 \times 10^6$ ($2.3 \times 10^6$). (d) DC Stark effect can be achieved by applying a DC electric field to the two-level system. The DC field blue-shifts one ring and red-shifts the other one. The resulting transmission spectra show an anti-crossing curve due to the optical coupling between the two rings. From the asymptote of the anti-crossing, a microwave-optical coupling efficiency of $g = 2\pi \times 0.5$ GHz/V is extracted. NT: normalized transmission.}
\end{figure*}

Electro-optic methods \cite{capmany_microwave_2007,xu_micrometre-scale_2005,ayata_high-speed_2017,wade_wavelength_2015,liu_graphene-based_2011,shambat_ultra-low_2011,shen_deep_2017} are ideally suited for controlling photonic two-level systems since they have fast response, can be programmed and allow for large scale integration.To realize a fully unitary photonic two-level system with coherent electro-optic control, the photon lifetime of each energy state needs to be much longer than the time required to drive the system from one state to the other. On the one hand, Large optical systems with optical amplifiers \cite{spreeuw_classical_1990} can emulate a classical two-level system but the quantum coherence of the optical photons is destroyed in the process. On the other hand, conventional integrated photonic platforms have not been able to meet the requirements of long photon lifetime and fast modulation simultaneously. For example, fast phase modulators  \cite{kues_-chip_2017,karpinski_bandwidth_2017} can generate new optical frequencies but they do not support distinct and long-lived optical modes. On the other hand, ultra-high \textit{Q} on-chip resonators have traditionally been realized in passive materials, such as silicon dioxide  (SiO$_2$) \cite{armani_ultra-high-q_2003,lee_chemically_2012} and silicon nitride (Si$_3$N$_4$) \cite{ji_ultra-low-loss_2017,bauters_ultra-low-loss_2011}, which can only be controlled electrically using slow thermal effect. Electrically active photonic platforms based on silicon \cite{xu_micrometre-scale_2005}, III-V materials \cite{shambat_ultra-low_2011,guo_-chip_2016}, plasmonics \cite{ayata_high-speed_2017}, graphene \cite{liu_graphene-based_2011}, and polymers \cite{enami_hybrid_2007} allow for fast electro-optic modulation at gigahertz frequencies, but suffer from dramatically reduced photon lifetimes compared to passive platforms. 

In this work, we overcome the existing performance trade-off paradigm and realize a programmable photonic two-level system that can be dynamically controlled using gigahertz microwave signals (Fig. \ref{fig1}a). Specifically, we create a microwave addressable photonic molecule using a pair of $80$-$\mu m$ radius integrated lithium niobate (LN) microring resonators patterned in close proximity to each other (Fig. \ref{fig1}b). The low optical loss \cite{zhang_monolithic_2017} and the efficient co-integration microwave electrodes \cite{guarino_electrooptically_2007,witmer_high-q_2017} allow us to simultaneously achieve large electrical bandwidth ($>$ 30 GHz), strong modulation efficiency (0.5 GHz/V) and long photon lifetime (2 ns). 

The photonic molecule supports a pair of well-defined optical energy levels, which are evident from the optical transmission measured using a tuneable telecom wavelength laser (Fig. \ref{fig1}c; also, see Supplementary for full spectrum). The two optical energy levels are formed by the evanescent coupling of light from one resonator to another through a 500-nm gap. When the optical coupling strength $\mu$ exceeds the optical loss rate $\gamma$ of each cavity, the coupling leads to a normal mode splitting resulting in a frequency doublet consisting of a lower frequency symmetric (S) and a higher frequency antisymmetric (AS) optical mode (Fig. \ref{fig1}a and c). The S (AS) mode spatially spans both optical cavities, with the light in the two cavities being in- (out-of-) phase. The two new eigenmodes, separated in frequency by the optical coupling strength, are the energy states of the photonic molecule. In our case, the two optical modes are separated by $2\mu = 2\pi\times 7$ GHz and have cavity linewidths of $\gamma = 2\pi \times 96$ MHz, corresponding to a loaded (intrinsic) quality factors of $Q = 1.9 \times 10^6$ ($Q_i = 2.5 \times 10^6$), thus forming a well-resolved two-level system (Fig. \ref{fig1}c). 

\begin{figure*}
	\centering
	\includegraphics[angle=0,width=\textwidth]{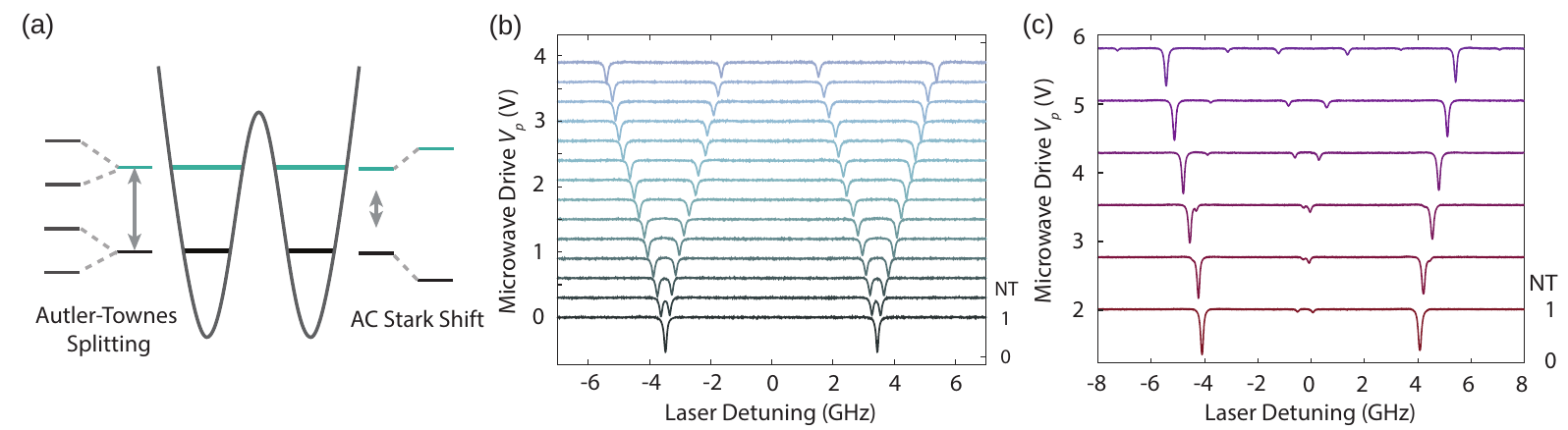}
	
	\caption{\label{fig2}\textbf{Microwave dressed photonic states.} (a) When the applied microwave frequency is tuned to the match the mode separation, dissipative coupling leads the two photonic levels to split into four levels - the effect referred to as the microwave induced photonic Autler-Townes (AT) splitting.  When the microwave is detuned far from the photonic mode splitting, the photonic energy levels experience a dispersive effect, leading to photonic AC Stark shifts. (b) Measured AT splitting in the photonic molecule where the splitting can be accurately controlled by the amplitude of the applied microwave. (c) Measured photonic AC Stark shifts for a microwave signal at 4.5 GHz. NT: normalized transmission.}
	
\end{figure*}

We induce photonic transitions in the two-level system using high-frequency electro-optic phase modulation of the two modes. The phase modulation is realized through the Pockels effect ($\chi^{(2)}$) of LN, where the optical refractive index can be changed by an applied electric field, with a response time on the femtosecond scale. To enable strong overlap between microwave and optical fields without significantly increasing the optical loss, we place gold microelectrodes 2.5 $\mu m$ away from the edge of the rings that form the photonic molecule (Fig. \ref{fig1}b). Importantly, the microwave circuit layout is designed to induce a phase delay on one ring and a phase advance on the other ring therefore introducing coupling between the spatially orthogonal S and AS modes (Fig. \ref{fig1}a and Supplementary).

We explore the analogy between an atomic two-level system and the photonic two-level system, and leverage it to demonstrate the control of the photonic molecule. We note that in our system the electro-optic effect plays the role equivalent to that of an electric dipole moment in the case of an atomic two-level systems, where in both systems external electromagnetic fields are used to couple and address the energy levels. Building on this analogy, we show that the energy levels of the photonic molecule can be programmed by applying a direct current (DC) electric field, the effect equivalent to a DC Stark shift used to control the energy levels in an atom. This is accomplished by applying a DC bias voltage in the range of $\pm$ 15 V to the microwave electrodes, which results in an avoided crossing curve shown in figure \ref{fig1}d. The extracted tuning/modulation efficiency $g = 2\pi\times 0.5$ GHz/V is an order of magnitude larger than previously demonstrated in bulk electro-optic resonator systems \cite{rueda_efficient_2016,savchenkov_tunable_2009}, and is due to the highly efficient overlap between microwave and optical fields enabled by our system \cite{wang_nanophotonic_2018}.

Next, we use a continuous wave (CW) coherent microwave field to control our photonic two-level system. This situation is similar to an atomic two-level system under a strong coherent excitation, with an important difference that in our case the number of photons that could populate each of the two levels is not limited to one, resembling that of an atomic ensemble. When the microwave frequency matches the energy difference of the two levels, an effective coupling between the two initially decoupled S and AS modes is introduced, leading to microwave-induced photonic Autler-Townes splitting (Rabi splitting), shown in Fig. \ref{fig2}a. This is similar to the emergence of dressed states in the case of an atomic two-level system resonantly excited with CW light. In our system, the splitting frequency ($\Omega$) can be precisely controlled up to several gigahertz by controlling the amplitude of the microwave signals (Fig. \ref{fig2}b). When the microwave frequency is far detuned from the transition frequency, an effective dispersive effect in the level splitting is induced, analogous to the AC Stark shift in atomic systems (Fig. \ref{fig2}c). Importantly, this effect can be used to control the effective coupling strength between the energy levels of the photonic molecule, which are otherwise determined by geometric factors.

\begin{figure*}[t!]
	\centering
	\includegraphics[angle=0,width=0.7\textwidth]{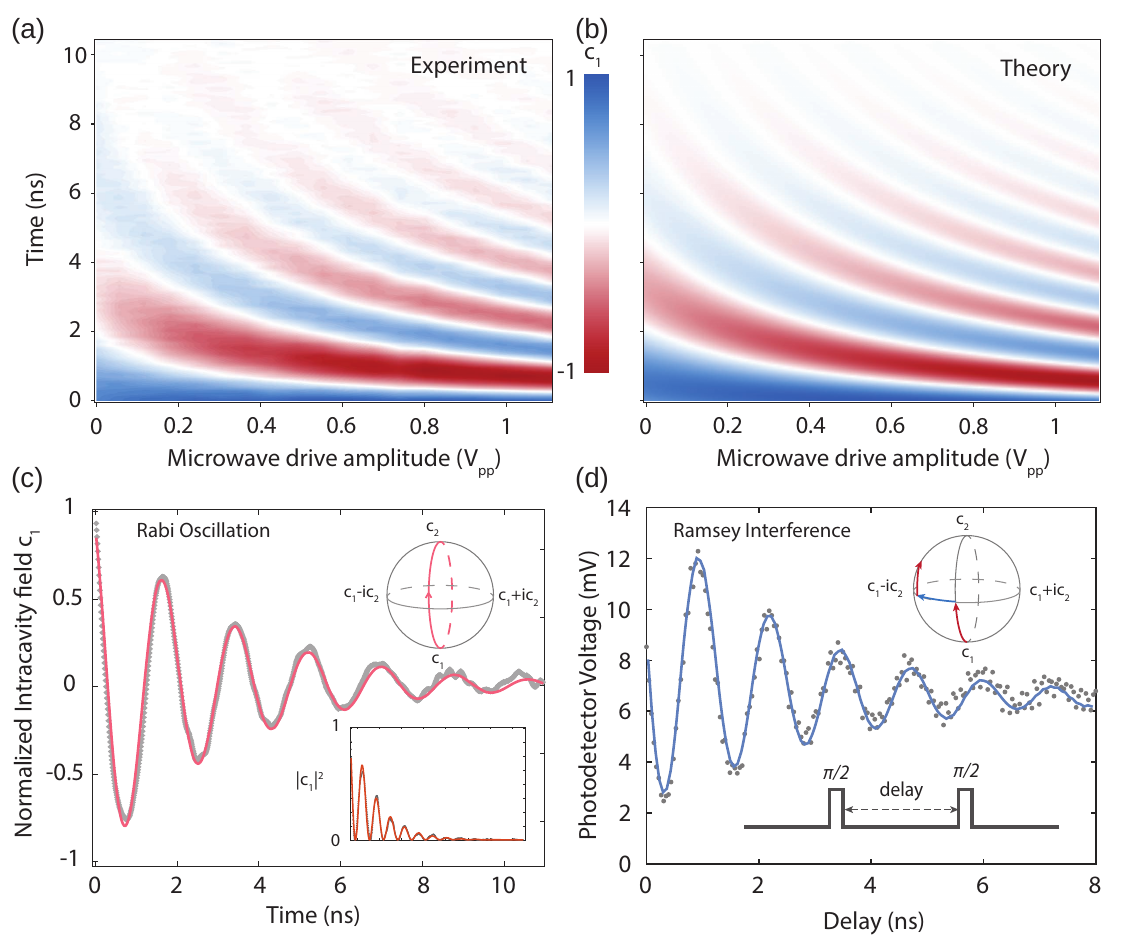}
	
	\caption{\label{fig3}\textbf{Coherent spectral dynamics in the photonic molecule} (a-b), Measured (a) and theoretically predicted (b) Rabi oscillation for different microwave strengths applied to the photonic two-level system indicating the control of the Rabi frequency over a wide range. The excitation laser is tuned to the frequency of the symmetric mode (i.e. mode $c_1$). As the microwave field is turned on at $t = 0$, the photon oscillates between the two energy levels leading to the observed signals. The measurements are in excellent agreement with theoretical predictions (see Supplementary). (c) Rabi oscillation observed for microwave $V_p = 1.1 V$ at $1.1$ GHz. This corresponds to rotation along the real-axis of the Bloch sphere. Inset shows the corresponding photon population (measured) in mode $c_1$.  (d) Oscillatory signals on a photodetector showing photonic Ramsey interference observed when $\delta$-detuned microwave $\pi/2$pulses are applied. The first $\pi/2$ pulse prepares the light in a superposition of the two optical modes with a precession frequency determined by $\delta$. The second $\pi/2$ pulse with a delay $\tau$have oscillating projections on to $c_1$ resulting in the measured Ramsey fringes. The solid curve shows calculated signal (see Supplementary). The optical coherence time is $\sim 1.6$ ns for both Rabi and Ramsey oscillations in agreement with the cavity decay time.}
\end{figure*}

We demonstrate the photonic molecule can be used for unitary transformation of light in the frequency domain using Rabi oscillation and Ramsey interference (see Supplementary). The observed Rabi oscillation corresponds to a rotation along the real axis of the energy Bloch sphere (Fig. \ref{fig3}c, inset) and indicates that light tunnels back and forth between the two optical modes at two different optical frequencies. In other words, using the language of nonlinear optics, applied microwave signal drives a sequence of resonance enhanced sum- and difference-frequency generation (SFG and DFG) processes that result in energy of photons being changed several times (more than 10 in Fig. \ref{fig3}c) before it is eventually dissipated due to the cavity photon loss (cavity life time $\sim$ 1.6 ns). To study the Rabi oscillation, we initialize the system by coupling a CW laser into the S mode and measure the real-time optical transmission as the microwave drive is turned on (see Methods and Supplementary). We find a large range of Rabi frequencies achievable with low applied voltages (Fig. \ref{fig3}a) to be in excellent agreement with theoretical predictions (Fig. \ref{fig3}b and Supplementary). In particular, for a peak drive voltage of $V_p = 1.1$ V, we observe coherent Rabi oscillation with a frequency of 1.1 GHz, being $ 16\%$ of the initial mode splitting. Even stronger driving regimes, where the Rabi frequencies are close or even exceed the level splitting, could be accessible in our system, thus enabling exploration of the extreme conditions where the rotating-wave approximation completely breaks down \cite{casanova_deep_2010}.

\begin{figure*}
\centering
\includegraphics[angle=0,width=0.8\textwidth]{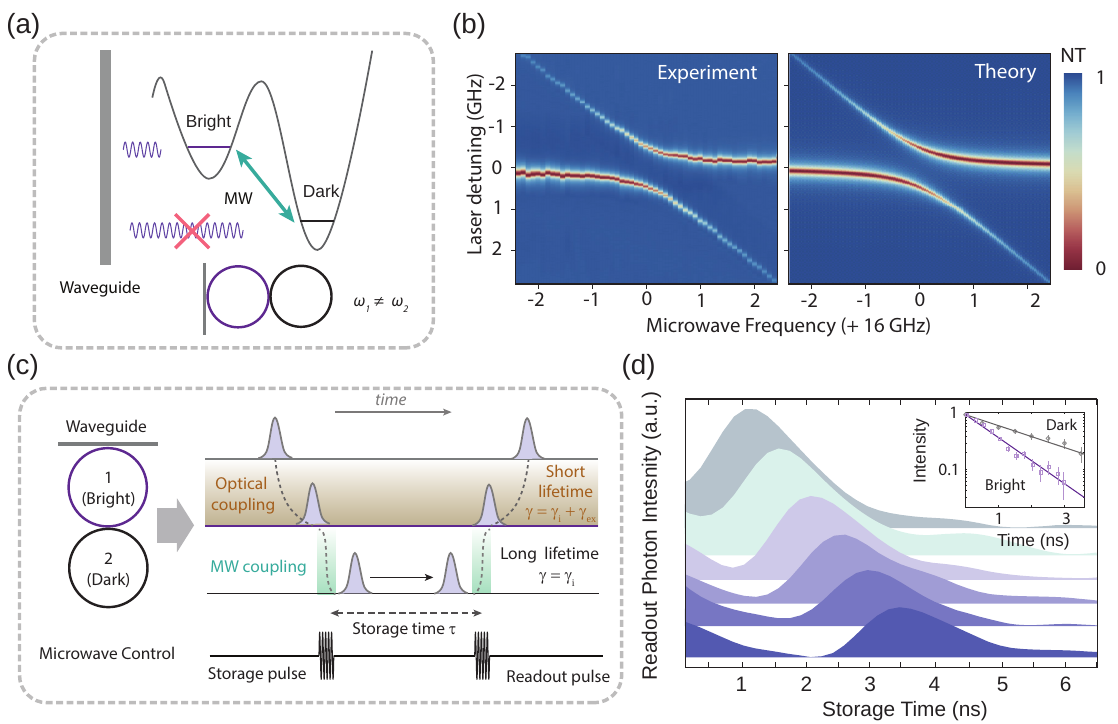}
\caption{\label{fig4}\textbf{On-demand storage and retrieval of light using a photonic dark mode} (a) The photonic molecule is programmed so that the resonance frequency of ring 1 is different from ring 2, resulting in bright and dark modes. The former is localized mainly in ring 1 while the latter in ring 2. As a result, bright mode can be accessed from the optical waveguide, while the dark mode cannot (forbidden by geometry). (b) A microwave field applied to the system can induce an effective coupling between bright and dark modes, indicated by the avoided crossing observed in the optical transmission spectrum. (c) Light can be deterministically stored and retrieved using the bright-dark mode pair and microwave control. Photons are coupled to ring 1 through fixed waveguide coupling. A microwave $\pi$pulse is applied to temporarily open a channel that transfers light from the bright to the dark mode. As the microwave is turned off, photons are restricted from any external waveguide coupling. After a certain desired storage time, a 2nd microwave $\pi$pulse retrieves the photon from the dark to the bright mode and the light is transferred back to the optical wave guide. $\gamma$ is the decay rate of the bright optical mode, $\gamma_\textrm{i}$ and $\gamma_\textrm{ex}$ are the intrinsic damping and waveguide coupling rate respectively. (d) The retrieved light from the dark mode measured at different time delays. Inset: the extracted intensity of the retrieved light shows nearly twice the lifetime of the critically coupled bright mode. This is expected since the lifetime of the bright mode is affected by the waveguide coupling, while the lifetime of the dark mode is determined by the intrinsic losses of the resonator only, as it is decoupled from the input waveguide. NT: normalized transmission.}
\end{figure*}

To show the control over the relative phase of the two photonic levels in our system (i.e. rotation along the imaginary axis), we measure photonic Ramsey interference by driving the system with detuned microwave pulses at 7.8 GHz. After initializing the system in the S mode, we apply a microwave $\pi/2$ pulse to transform the optical field into a superposition of the two states. As the microwave is turned off for a time period of $\tau$, the superposition state precesses along the equator of the Bloch sphere at a rate determined by the microwave detuning $\delta$ After sending another microwave $\pi/2$ pulse, the photons are rotated to different energy states, determined by the delay $\tau$, and measured using a photodetector. The result is the so-called Ramsey interference shown in figure \ref{fig3}d. As the optical lifetime of the two-level system ($\sim$ 1.6 ns) is much shorter than the coherence time of the laser ($\sim \mu$s), the phase coherence time of the two energy levels obtained in this measurement is dominated by cavity dissipation in good agreement with that obtained from the Rabi oscillation. 

We leverage the ability to perform unitary transformations in the frequency domain to achieve on-demand photon storage and retrieval - a critical task for optical signal processing. While a static resonator can be used to slow down the propagation of a photon, such slow-down is fundamentally limited by the frequency bandwidth of the resonator, i.e. the delay-bandwidth product, and cannot be controlled on-demand. In contrast, the use of a dynamically-modulated resonator system can overcome the delay-bandwidth product constraint, enabling new functionalities such as optical buffering \cite{yanik_stopping_2004}. To enable controllable write and read of photons into a resonator and from an external waveguide, the optical coupling strength needs to be altered faster than the cavity photon lifetime. To achieve this, we apply a large DC bias voltage (15 V) to reconfigure the photonic molecule into a pair of bright and dark modes. In this limit, one of the modes is mainly localized in the 1st ring (purple in Fig. \ref{fig4}a), and thus is still accessible to the input optical waveguide and is optically bright, while the other mode is mainly localized in the 2nd ring (black in Fig. \ref{fig4}a) and thus is decoupled from the input optical waveguide by geometry and becomes optically dark (See Supplementary for details). Notably, optical access to the dark optical mode can be granted by applying a microwave signal with frequency matched to the difference between the two optical modes (Fig. \ref{fig4}a). The microwave modulation results in an effective coupling between the bright and the dark mode, which we directly observe from the avoided crossing in the spectrum of the bright optical mode (Fig. \ref{fig4}b). In our experiment, the waveguide coupling is designed to be critically-coupled to a single resonator maximizing the transmission extinction ratio. We excite the critically-coupled bright mode from the optical waveguide, and then apply a microwave $\pi$ pulse to switch light from the bright to the dark mode (Fig.\ref{fig4}c). Once the microwave signal is turned off, the photons are trapped in the dark mode and become decoupled from the waveguide leaving cavity intrinsic dissipation as the only photon loss mechanism. After a desired storage time, we apply another microwave $\pi$ pulse to deterministically retrieve the photons from the dark mode back into the bright mode and then into the optical waveguide (Fig. \ref{fig4}c). Tracking the intensity of the retrieved optical pulses stored in the dark mode, we extract a dark mode lifetime of 2 ns, which is about two times of the lifetime of the critically coupled bright mode as expected (Fig. \ref{fig4}d). Using an over-coupled bright optical modes, while further improving the quality factor of the integrated lithium niobate resonators towards its material limit ($> 10^9$), could result in a tunable storage time of hundreds of nanoseconds (see Supplementary). 

Our demonstration of the coherent and dynamic control of a two-level photonic molecule with microwave fields and on-demand photon storage and retrieval paves a path to a new paradigm of control over photons. These results represent the initial step towards integrated electro-optic coherent manipulation of photonic states and energies, and could have immediate applications in signal processing and quantum photonics. With microwave control and the possible integration of on-chip photonic components including filters, routers and modulators, a new generation of photonic-electronic systems with advanced functionalities can be put in practice. Considering the high versatility and scalability of dynamically controlled two- and multi-level photonic systems, they have the potential to enable novel photonic technologies including topological photonics \cite{lin_photonic_2016}, advanced photonic computation concepts \cite{mcmahon_fully_2016,shen_deep_2017} and on-chip optical quantum systems \cite{obrien_optical_2007,kues_-chip_2017}. 

This work was supported in part by National Science Foundation Grant (ECCS1609549, DMR-1231319), Office of Naval Research MURI grant N00014-15-1-2761 and Army Research Laboratory Center for Distributed Quantum Information W911NF1520067, Center for Integrated Quantum Materials (CIQM). Device fabrication was performed at CNS at Harvard University.

\bibliography{reference}

\pagebreak
\onecolumngrid

\begin{center}
	\textbf{\large Supplemental Information}
\end{center}

\setcounter{equation}{0}
\setcounter{figure}{0}
\setcounter{table}{0}
\setcounter{page}{1}
\makeatletter
\renewcommand{\theequation}{S\arabic{equation}}
\renewcommand{\thefigure}{S\arabic{figure}}
\renewcommand{\bibnumfmt}[1]{[S#1]}
\renewcommand{\citenumfont}[1]{S#1}
\section{Summary of Fabrication and Measurements Methods}
Devices are fabricated on a single crystalline thin-film lithium niobate (LN) device layer bonded onto a silicon (Si) handle wafer with a 2 $\mu$m thick thermally grown silicon dioxide layer on top. Standard electron-beam (e-beam) lithography is used to realize optical waveguides and microresonator patterns in hydrogen silsequioxane (HSQ) e-beam resist. The patterns are then transferred into the LN layer using argon plasma etching in a standard inductively couple plasma reactive ion etching (ICP-RIE) tool. The etched depth is 350 nm leaving a 250 nm LN slab behind. The slab allows for efficient electrical field penetration into the waveguide core region. The first layer of the gold interconnects is patterned using e-beam lithography and the metals are deposited with e-beam evaporation methods and lift-off processes. Next, a 1.6 $\mu$m silicon dioxide layer is deposited on top using plasma enhanced physical vapour deposition (PECVD) method. Finally, metal vias and the top metal layer are realized using a standard photolithography followed by e-beam evaporation and lift-off processes.

The light from a tunable telecom wavelength laser (SANTEC TS510) is launched into, and collected from, the LN waveguides using a pair of lensed optical fibres. The microwave control signals are generated from an arbitrary waveform generator (AWG, TEKTRONIX 70001A), before they are sent to electrical amplifiers. Electrical circulator or isolators are used to prevent multiple electrical reflections. For the Rabi oscillation measurements, the electric field amplitude ($c_1$) in the S mode is measured by interfering the light out-coupled from the double-ring system with the pump light in the optical waveguide. The interference produces a homodyne signal for $c_1$ that is sent to a 12 GHz photodiode (Newport 1544A), and due to the optical frequency difference, the rapid interference signal between the pump light and $c_2$ can be filtered out electrically using a low-pass filter. For the Ramsey measurements, the optical power is sampled after the 2nd $\pi/2$ pulse using the fast photodiode. For the photon storage measurements, the pump light is synchronously turned off with the first $\pi/2$ pulse allowing for direct power readout of the retrieved light and prevent pump further leaking into the bright mode. The modulation on the pump signal is achieved by an external electro-optic modulator synchronized with the microwave control signals.

	\section{Device and Experimental Setup Details}

	\begin{figure}[ht!]
		\centering
		\includegraphics[angle=0,width=0.8\textwidth]{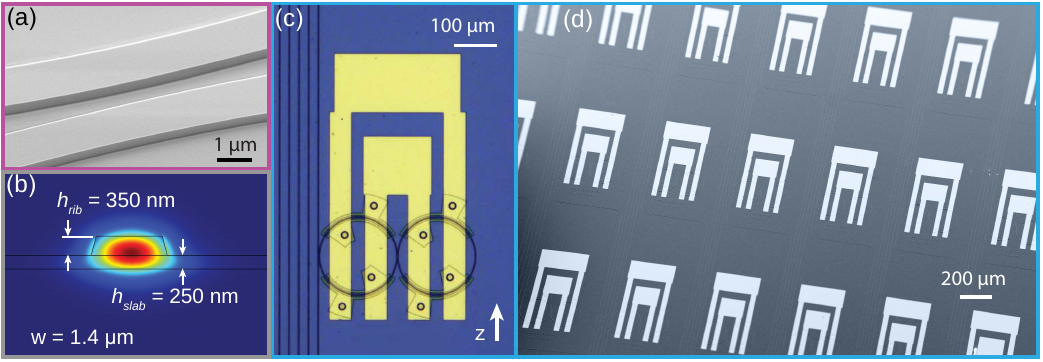}
		\caption{\label{figs2}\textbf{Device details.} (a) scanning electron microscope (SEM) image of the gap between the coupled microring resonators. (b) Cross-section of the optical mode profile in the ring resonator. (c) microscope image of the full device showing the double-ring and microwave electrodes. (d) SEM image of the array of double ring devices fabricated on a single chip.}
		\label{fig:devices}

	\end{figure}
	
	The ring resonators of the photonic molecule have waveguide width of 1.4 $\mu$m and vary coupling gaps of $\sim 700$ nm. The coupling waveguides are 800 nm wide by 600 nm thick with a rib height of 350 nm and slab height of 250 nm. This ratio is chosen to ensure optimum electro-optic over-lap while still maintaining a tight bending radius.
	
	The high frequency microwave electrodes are designed to achieve differential driving of the two resonators. As shown in figure \ref{fig:devices}, the top contact pad is connected to the bottom electrodes of the left ring and the top electrodes of the right ring (and vise versa for the bottom contact pad). Since the LN thin film is $x$-cut (extraordinary axis $z$ in plane), such configuration allows the electrical field to be in the same direction on each ring, while opposite for the two rings. Here we pattern 60 programmable photonic molecules on a 10 mm by 8 mm thin film lithium niobate (LN) chip.
	
	\begin{figure}[ht!]
		\centering
		\includegraphics[angle=0,width=0.8\textwidth]{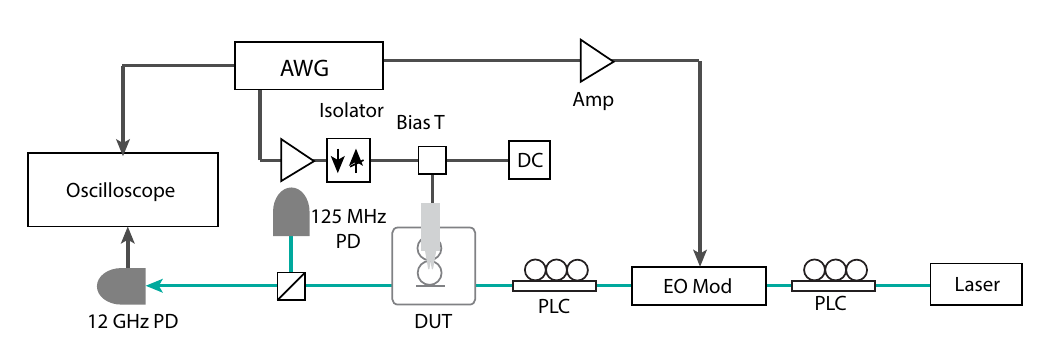}
		
		\caption{\label{figs1}\textbf{Experimental Setup.} The device is optically pumped by a tunable telecom laser centered around 1630 nm. The light is sent through an external electro-optic modulator and polarization controllers before coupling into the chip with a lensed fiber. The output optical signal, also coupled with a lensed fiber, is sent to a 12 GHz photodetector (New focus 1811A). The converted electrical signal is directed to an oscilloscope. The microwave control signals are generated by an arbitrary wave generator (AWG, Tektronix 70001A) and amplified before sent on to the device. A bias T is used to allow DC control on the microresonators. An electrical isolator is used to capture the electrical reflection from the microresonators. The oscilloscope, device drive signals and modulator drive signals are all synchronized.}
		
	\end{figure}
	
	\begin{figure}[ht!]
		\centering
		\includegraphics[angle=0,width=0.9\textwidth]{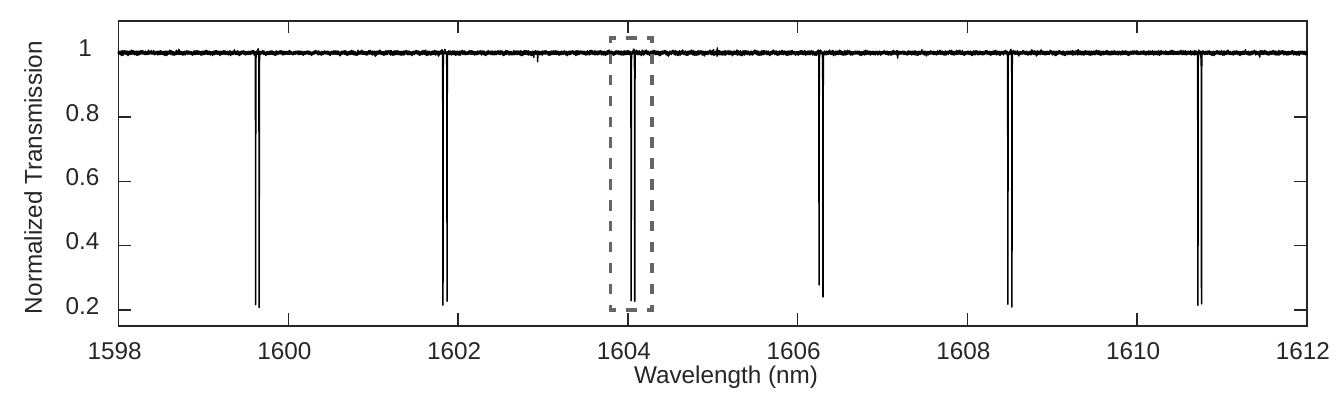}
		
		\caption{\label{figs3}\textbf{Measured transmission spectrum.} Typical normalized transmission spectra of a double-ring resonator showing pairs of photonic two-level doublets, separated by $\sim$ 2.2 nm of free-spectral-range.}
		\label{fig:trans}
	\end{figure}
	The optical properties of the devices are characterized using a tunable telecom laser (Santec TSL-550). The polarization of the light is tuned to TE (in-plane of the chip) before sending into the chip with a lensed fiber. The DC transmission is measured on a 125 MHz photodetector (New Focus 1811A) and the fast oscillation signal is measured on a 12 GHz photodetector (New Focus 1544A) and a fast real-time oscilloscope (Tektronix MSO71604C). Figure \ref{fig:trans} shows the normalized transmission spectrum of the photonic molecule, showing pairs of two-level resonances separated by the free-spectral-range of the microring resonator ($\sim$ 2.2 nm). 
	
	The microwave signals are prepared using an arbitrary waveform generator (Tektronix 70001A), which provides synchronized signals to drive the photonic molecule as well as an external modulator and the oscilloscope. The microwave driving signal for the photonic molecule is amplified and sent through an isolator to minimize reflection to the amplifier. A bias-T is used to coupled DC field into the resonator for fine tuning.

	\section{Description of the photonic molecule}
	
	\subsection{System Hamiltonian}
	The nanophotonic two-level system under a coherent microwave drive can be described by a Hamiltonian with the form
	\begin{equation}
	\label{eq:ham}
	\hat{H}=\omega_1 a_1^\dagger a_1+\omega_2 a_2^\dagger a_2 + \mu (a_1^\dagger a_2+a_1 a_2^\dagger) +\Omega (a_1^\dagger a_1-a_2 a_2^\dagger) \textrm{cos} (\omega_m t)
	\end{equation}
	where $a_1$ ($a_1^\dagger$) and $a_2$ ($a_1^\dagger$) are the annihilation (creation) operators of the two optical modes of the respective microresonator, $\mu$ is the coupling strength between the two optical resonators, $\Omega=g V_0$ is the interaction strength of microwave field to the optical resonator, determined by the coupling strength $g$ and microwave peak amplitude $V_0$. $\omega_m$ is the frequency of the microwave modulation.
	
	\subsection{Identical resonator: two bright modes}
	When the resonant frequencies of the two resonators are identical, i.e. $\omega_1 = \omega_2 \equiv \omega_0$, equation \eqref{eq:ham} can be expressed as
	
	\begin{equation}
	\label{eq:hamid}
	\hat{H}=\omega_+ c_1^\dagger c_1+ \omega_- c_2^\dagger c_2 +\Omega (c_1^\dagger c_2+c_1 c_2^\dagger) \textrm{cos} (\omega_m t)
	\end{equation}
	
	where  $\omega_{\pm}=\omega_{0}\pm \mu$,  $c_1^{(\dagger)}=\frac{1}{\sqrt{2}}(a_1^{(\dagger)}+a_2^{(\dagger)})$ and $c_2^{(\dagger)}=\frac{1}{\sqrt{2}}(a_1^{(\dagger)}-a_2^{(\dagger)})$.
	
	To simplify the system, we apply a unitary transformation $U_1=e^{(i \omega_+ c_1^{\dagger} c_1 t + \omega_- c_2^{\dagger} c_2 t)}$ and the rotating wave approximation (RWA), we obtain
	
	\begin{equation}
	U_1\hat{H}U_1^{\dagger}= \frac{\Omega}{2} (c_1^\dagger c_2 e^{i\delta t}+c_1 c_2^\dagger e^{-i \delta t})
	\label{eq:hmit}
	\end{equation}
	
	where the microwave detuning $\delta=\omega_m-\Delta \omega$ and $\Delta\omega$ is the frequency difference between $c_1$ and $c_2$. Equation \ref{eq:hmit} is equivalent to the time independent Hamiltonian
	
	\begin{equation}
	\label{eq:hamid_1}
	\hat{H}=\frac{\delta}{2} c_1^\dagger c_1 - \frac{\delta}{2} c_2^\dagger c_2 +\frac{\Omega}{2} (c_1^\dagger c_2+c_1 c_2^\dagger)
	\end{equation}

	We obtain the following equations of motion using input-output theory from equation \ref{eq:hamid_1},
	\begin{equation}
	\frac{\d}{\d t}
	\begin{pmatrix}
	c_1 \\
	c_2
	\end{pmatrix}
	=
	\begin{pmatrix}
	-i \frac{\delta}{2} & - i \frac{\Omega}{2}\\
	- i \frac{\Omega}{2} &i \frac{\delta}{2}
	\end{pmatrix}
	\begin{pmatrix}
	c_1 \\
	c_2
	\end{pmatrix}
	\label{eq:eom}
	\end{equation}

	\subsection{Autler-Townes Splitting and AC Stark Shifts}
	
	\begin{figure}[ht!]
		\centering
		\includegraphics[angle=0,width=0.8\textwidth]{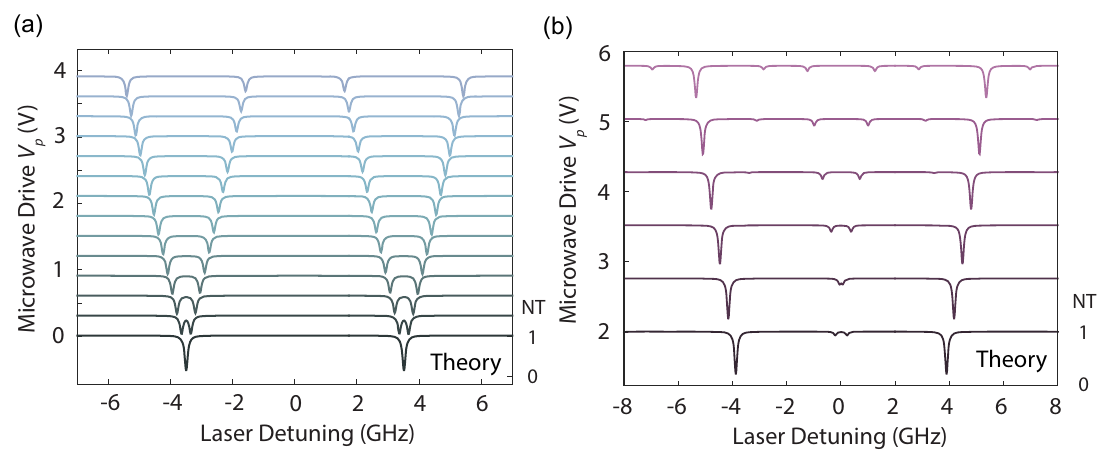}
		
		\caption{\label{figs4}\textbf{Calculated optical transmission spectrum for Autler-Townes (AT) splitting and AC Stark shift} (a) Simulated AT spectrum with full mode described in equation \ref{eq:hamid}. (b) Simulated AC Stark shift spectrum with full mode described in equation \ref{eq:hamid}}
		\label{fig:at}
	\end{figure}
	
	Under a continuous wave microwave modulation with frequency close to the optical two-mode splitting, each resonance of $c_1,c_2$ will further split into two dips with splitting frequency
	\begin{equation}
	W=\sqrt{(\Omega^2+\delta^2)}
	\label{eq:at}
	\end{equation}
	obtained by diagonalizing the coupling matrix
	$M=\begin{pmatrix}
	-i \delta/2 & - i \Omega/2\\
	- i \Omega/2 &i \delta/2
	\end{pmatrix}$ from equation \ref{eq:eom}.
	
	For the Autler-Townes splitting measurement data shown in Fig. \ref{fig2}a in the main text and simulation shown in Fig. \ref{fig:at}, the microwave modulation signal is on-resonance with the optical two-mode splitting, i.e. $\delta = 0$, with $W=\Omega=g V_0$.
	
	When the microwave frequency detuning to the two-mode resonances is much larger than the driving strength (i.e. $\delta\gg\Omega$), the microwave modulation behaves as a weak perturbation. This perturbation induces a shift of the optical energy levels. Applying the second-order perturbation theory, we have 
	\[M
	\approx
	\begin{pmatrix}
	-i \left(\frac{\delta}{2}+\frac{\Omega^2}{4\delta}\right) & \\
	& -i \left(-\frac{\delta}{2}-\frac{\Omega^2}{4\delta}\right)
	\end{pmatrix}
	\]
	
	In the static frame we obtain the shifted frequency:
	\[
	\omega_\pm^\prime=\omega_\pm \mp\frac{\Omega^2}{4\delta}
	\]
	In the experimental measurements in Fig. \ref{fig2}b and \ref{fig:at} we have $\delta<0$ where the two energy levels of $c_1,c_2$ moves away from each other. For a blue-detuned microwave drive signal, i.e. $\delta>0$, the two energy levels will move towards to each other.
	
	\subsection{Optical two-level dynamics}
	To study the dynamics of the photonic two-level system, we consider the equations of motion \ref{eq:eom} with optical loss and laser input fields. For a single tone optical input on resonance with $c_1$, the equations of motion are:
	
	\begin{align}
	&\dot{c_1}=\left(-i \frac{\delta}{2}-\frac{\gamma}{2}\right) c_1 - i \frac{\Omega}{2} c_2
	-\sqrt{\gamma_{\textrm{ex}}/2}s_{\textrm{in}} \label{eq:motion1}
	\\
	&\dot{c_2}=\left(
	i \frac{\delta}{2}
	-\frac{\gamma}{2}\right) c_2 - i \frac{\Omega}{2} c_1
	-\sqrt{\gamma_{\textrm{ex}}/2}
	s_{\textrm{in}}e^{\i \omega_m t}
	\label{eq:motion2}
	\end{align}
	where $\gamma$ is the total decay rate of the modes $c_{1,2}$, $\gamma_\textrm{ex}$ is the waveguide coupling rate to mode $a_1$ and $s_\textrm{in}$ is the input laser field. After dropping the fast rotating terms with frequency of $\omega_m$, we obtain the dynamical solution of $c_1$ and $c_2$,
	
	\begin{equation}
	\begin{aligned}
	c_1&=c_{10}e^{-\frac{\gamma}{2}t}  \left( \cos \frac{W}{2}t-\i \frac{\delta}{W}\sin \frac{W}{2}t\right)\\
	c_2&=-\i c_{10} e^{-\frac{\gamma}{2}t}\frac{\Omega}{W} \textrm{sin}\frac{W}{2}t
	\label{eq:sol1}
	\end{aligned}
	\end{equation}

	where $c_{10}=\frac{\sqrt{\gamma_\textrm{ex}} s_\textrm{in}}{-i\Delta-\gamma/2}$ is the steady state amplitude in mode $c_1$ before the microwave is turned on and $\Delta$ is the laser detuning. Here $W=\sqrt{\Omega^2+\delta^2}$ is the Autler-Townes splitting frequency as in equation \ref{eq:at} and is also the Rabi oscillation frequency. The output power from the waveguide is then
	
	\begin{equation}
	\begin{aligned}
	P_{\textrm{out}}&=|s+\sqrt{\gamma_{\textrm{ex}}}a_1|^2\\
	&=|s+\sqrt{\gamma_{\textrm{ex}}/2}(c_1+c_2 e^{i \omega_m t})|^2
	\label{eq:sol1out}
	\end{aligned}
	\end{equation}
	
\begin{figure}[ht!]
		\centering
		\includegraphics[angle=0,width=0.8\textwidth]{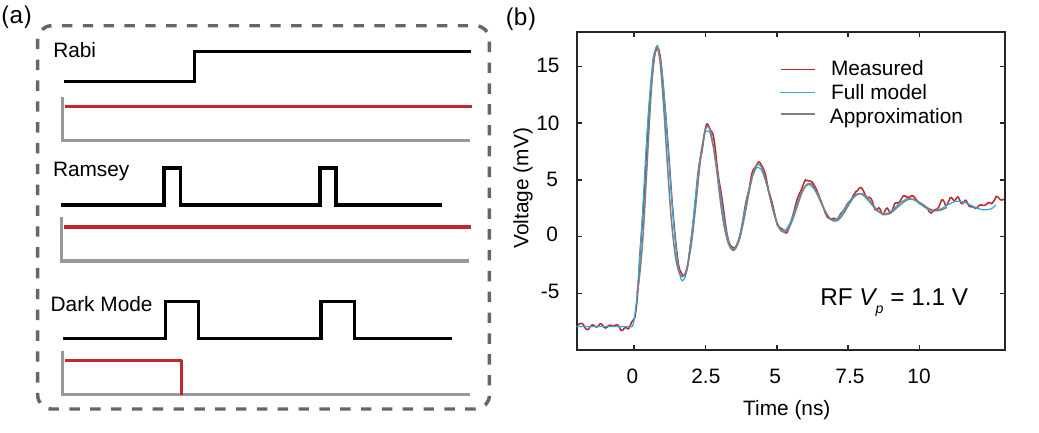}
		
		\caption{\label{figs5}\textbf{Rabi oscillation, Ramsey interference and dark mode measurements} (a)  Rabi oscillation is measured in real-time as the microwave signal is turned on. The laser entering the cavity after the turn-on of the microwave signal is negligible as the split optical mode is transparent at the original laser wavelength. Ramsey interference is measured by two time-delayed microwave $\pi/2$ pulses. The optical power is recorded at the end of the second pulse. For the photonic dark mode measurements, the laser is turned off by an external electro-optic modulator for direct detection of optical powers and prevent further coupling into the bright mode during the delay time. (b) Comparison of the measured photodetector voltage, the simulated transmission of the full system and the calculated output from equation \ref{eq:sol1app}}
		\label{fig:dyn}
		
	\end{figure}

	Substitute equations \ref{eq:sol1} into the expression of the output power and set $\delta=0$, we obtain
	
	\begin{equation}
	\begin{aligned}
	P_{\textrm{out}}=P_\textrm{in}\left|1-\frac{2\gamma_{ex}}{\gamma}e^{-\frac{\gamma}{2}t}\left(
	\cos \frac{W}{2}t
	\right)
	\right|^2
	\label{eq:sol1app}
	\end{aligned}
	\end{equation}
	
	where $P_\textrm{in}=|s|^2$ and we made the approximation to drop the fast rotating terms of $c_2$ at $\sim 7$ GHz, as we restrict our measurement bandwidth to $\sim 2$ GHz. The intra-cavity field can therefore be directly calculated from the output intensity by $c_1=(1-\sqrt{P_{\textrm{out}}/P_\textrm{in}})\times \gamma/2\gamma_{ex}$. Therefore the real-time AC signal on the photodetector is directly related to the intracavity field $c_1$. Figure \ref{fig:dyn} shows the full numerical solution as of equation \ref{eq:hamid}, the approximated solution \ref{eq:sol1app} and the measured photodetector voltage.
	
	The Ramsey interference is achieved by applying two microwave $\pi/2$ pulses separated by a time delay $\tau$. Here the microwave signal is detuned from the level-splitting by $\delta$. Using the protocol of Ramsey interference and equation \ref{eq:motion1}, \ref{eq:motion2} , we can get the solution after two $\pi/2$ pulse as a function of time delay $\tau$:
	
	\begin{equation}
	\begin{split}
	c_1(2T+\tau)=e^{-\frac{\gamma}{2}(2T+\tau)}e^{-\i \frac{\delta}{2}\tau}
	\left[
	\frac{\Omega^2}{2W^2}(1-e^{\i \delta \tau})-\i \frac{\delta}{W}
	\right]c_{10}
	+e^{-\frac{\gamma}{2}T}e^{-\i \frac{\delta}{2}\tau}\left[1-e^{-\frac{\gamma}{2}\tau}\right]
	\left[
	\frac{1}{\sqrt 2}-\i \frac{\delta}{W} \frac{1}{\sqrt 2}
	\right]c_{10}
	\end{split}
	\label{eq:rmc1}
	\end{equation}
	
	\begin{equation}
	\begin{split}
	c_2(2T+\tau)=e^{-\frac{\gamma}{2}(2T+\tau)}e^{-\i \frac{\delta}{2}\tau}
	\left[
	\frac{\delta \Omega}{2W^2}(e^{\i \delta \tau}-1)-\i \frac{\Omega}{2W}(e^{\i \delta \tau}+2)
	\right]c_{10}
	+e^{-\frac{\gamma}{2}T}e^{-\i \frac{\delta}{2}\tau}\left[1-e^{-\frac{\gamma}{2}\tau}\right](-\i \frac{\Omega}{W}c_{10})
	\end{split}
	\label{eq:rmc2}
	\end{equation}
	
	where  from mode $c_1$. $T$ is the duration of the $\pi/2$ pulse. The first terms in equations \ref{eq:rmc1} and \ref{eq:rmc2} represent the intra-cavity fields exist in the cavity prior to the arrival of the first $\pi/2$ pulse. The second terms are the leakage of the pump signal into the cavity during the delay, which contributes to a DC term. So the oscillation of the output signal  $P_\textrm{out}=\left|s+\sqrt{\gamma_{ex}/2}(c_1+c_2 e^{i \omega_m t})\right|^2$ rotating at with the frequency $\delta$ corresponds to the population of $c_1$ after dropping the fast rotating terms related to $c_2$.
	
	\subsection{Optical dark modes}

	In the case where $a_1$ and $a_2$ are far detuned by $\delta \omega\gg \mu$, as in the photon storage and retrieval measurements, we prefer to eliminate the coupling term $\mu(a_1^\ad a_2+a_1 a_2^\ad)$ by a Bogoliubov transformation. Assuming a new basis of $c_1, c_2$ satisfying:
	\[
	c_1=va_2-ua_1
	\]
	\[
	c_2=ua_2+va_1
	\]
	
	Since $c_1,c_2$ needs to satisfy the bosonic commutation relationship, we have the condition $u^2+v^2=1$. So we set $u=\cos {\frac \theta 2}, v=\sin{\frac \theta 2}$ with $\tan\theta=\mu/\delta \omega$. Then result of this transformation gives us a Hamiltonian for $c_1,c_2$:
	\[
	\begin{split}
	\hat{H}=&  \omega_1 c_1^\ad c_1 +\omega_2 c_2^\ad c_2 +\Omega \cos (\omega_m t) \sin\theta (c_1 c_2^\ad+c_1^\ad c_2)\\&+ \Omega \cos (\omega_m t) \cos\theta (c_2^\ad c_2-c_1^\ad c_1)
	\end{split}
	\]
	where $\omega_1=\omega_0-\sqrt{\mu^2+\delta \omega^2},\omega_2=\omega_0+\sqrt{\mu^2+\delta \omega^2}$. This Hamiltonian indicate that for $c_1,c_2$, the microwave modulation has a component that induces an exchange interaction $c_1 c_2^\ad+c_1^\ad c_2$ and a component that induces a frequency modulation $c_2^\ad c_2-c_1^\ad c_1$.
	
	As for the bright mode pair case we discussed above, i.e. for a small bias $\mu\gg\delta \omega$, $a_1,a_2$ are nearly degenerate and in $c_1, c_2$ basis,
	\[
	c_1\approx -\frac {1} {\sqrt{2}} (a_1+a_2)
	\]
	
	\[
	c_2\approx \frac {1}{ \sqrt{2}} (a_2-a_1)
	\]
	
	whereas in the case of a birght-dark optical mode pair, i.e. the bias voltage is high $\mu \ll \delta \omega$, $a_1,a_2$ have a large frequency difference and in $c_1,c_2$ basis, $c_1$ is composed by a large part of $a_1$ with small part of $a_2$ and $c_2$ has large part of $a_2$ while has only small part of $a_1$. That means:
	\[
	c_1\approx -\left(a_1+\frac{|\mu|}{2\delta \omega}a_2\right)
	\]
	
	\[
	c_2\approx \left(a_2-\frac{|\mu|}{2\delta \omega}a_1\right)
	\]

	Here the conversion term is finite and is suppressed by a factor $\sin \theta=\frac{\mu}{\delta \omega}$ and the term $\Omega \cos (\omega_m t) \cos\theta (c_2^\ad c_2-c_1^\ad c_1)$ is large, meaning that $c_1,c_2$ are being frequency modulated. However, Since the modulation frequency is orders of magnitudes larger than the bandwidth of the optical modes, we can neglect this term under high-$Q$ approximation. The resulting Hamiltonian still has a similar form to the bright mode pairs, with a  pre-factor $\sin{\theta}$ in conversion efficiency.
	
\begin{equation}
	\begin{split}
	\hat{H}_\textrm{dark}=& \omega_1 c_1^\ad c_1+\omega_2 c_2^\ad c_2  +\Omega \cos (\omega_m t) \sin\theta (c_1 c_2^\ad+c_1^\ad c_2)
	\end{split}
	\label{eqn:dk}
\end{equation}
	
	Similar to the previous section, applying RWA and input-output theory to equation \ref{eqn:dk}, we obtain
	
	\[
	\begin{split}
	\dot c_1=&(-\i \omega_1 - \frac{\gamma_1}{2})c_1 -\i \frac{\Omega}{2} \sin \theta c_2- \sqrt{\gamma_{\textrm{ex1}}} s_\textrm{in} e^{-\i \omega_L t}
	\end{split}
	\]
	\[
	\begin{split}
	\dot c_2=&(-\i \omega_2 - \frac{\gamma_2}{2})c_2  -\i \frac{\Omega}{2} \sin \theta c_1- \sqrt{\gamma_{\textrm{ex2}}} s_\textrm{in} e^{-\i \omega_L t}
	\end{split}
	\]
	
	where $\gamma_{1,2}=\gamma_i+\gamma_\textrm{ex1,2}, \gamma_{\textrm{ex1}}=\gamma_\textrm{ex} \cos^2\frac \theta 2, \gamma_{\textrm{ex2}}=\gamma_\textrm{ex}\sin^2\frac \theta 2$, $\gamma_i$ is the internal loss of each ring and $\gamma_\textrm{ex}$ is the external loss of ring 1 to waveguide, and $\gamma_{ex1}, \gamma_{ex2}$ are the external losses of modes $c_1,c_2$ to waveguide. From these equations of motion, we see that at large bias voltages, $\gamma_{ex2}\rightarrow 0$ and the access of mode $c_2$ to the external waveguide is effectively controlled by the microwave drive $\Omega$. At the same time, the life time for mode $c_2$ becomes closer to that of the intrinsic cavity life time. Therefore, $\pi$-pulse control sequences can be implemented as described in the main text to achieve the on-demand photon-storage and retrieval. 
	
	The bright mode can be over-coupled to the optical waveguide while the dark mode is detuned far from the bright optical mode. As an example, assuming a lithium niboate ring resonator can be fabricated close to its material loss limit ($Q\sim10^9$), set $\delta\omega=20$ GHz, $\mu=1$ GHz so $\theta=0.05$, and let $\gamma_\textrm{ex1}=2$ GHz, one could achieve a tunable delay of more than 800 ns with a minimum of 500 ps steps (limited by output coupling bandwidth) and over $95\%$ efficiency.

\end{document}